\documentclass[twocolumn,showpacs,preprintnumbers,amssymb,aps,pra,]{revtex4}

\usepackage{amsmath}
\usepackage{graphicx}
\usepackage[dvips]{color}

\begin{document}

\title{Small Numbers of Vortices in Anisotropic Traps}

\author{S.~McEndoo}
\email{smcendoo@phys.ucc.ie}
\author{Th.~Busch}
\affiliation{ Department of Physics,
  National University of Ireland, UCC, Cork, Republic of Ireland}

\date{\today}

\begin{abstract}
  We investigate the appearance of vortices and vortex lattices in
  two-dimensional, anisotropic and rotating Bose-Einstein
  condensates. Once the anisotropy reaches a critical value, the
  positions of the vortex cores in the ground state are no longer given
  by an Abrikosov lattice geometry, but by a linear arrangement.
  Using a variational approach, we determine the critical stirring
  frequency for a single vortex as well as the equilibrium positions
  of a small number of vortices.
\end{abstract}




\maketitle

\section{Introduction}

Quantized vortices are one of the hallmarks of superfluidity and the
first observation of vortex lattices in liquid $^4$He
\cite{Yarmchuk:79}, and more recently in gaseous Bose-Einstein
condensates \cite{AboShaeer:01}, has lead to a large interest in
studying their ground states and dynamical properties. Gaseous
Bose-Einstein condensates are particularly well suited to the study of
vortices, as their internal and external system parameters, such as
the inter-atomic scattering length or the density, can be
experimentally controlled. This allows access to study a large
range of vortex states in different parameter regimes.

The structure of a ground state vortex lattice is given by the demand
to minimise the energy of the system, under the condition of fixed
angular momentum. For large angular momentum this is best done by
forming an Abrikosov (triangular) lattice of charge one
vortices~\cite{AboShaeer:01, Anglin:02}. In different parameter
regimes, however, other lattice structures have been shown to appear:
symmetry preserving annular arrangements for small
condensates~\cite{Romanovsky:08}, rectangular lattices in traps with
weak, superposed optical lattices \cite{Reijnders:04}, giant vortices
in combined harmonic and quartic traps \cite{Aftalion:04}, and linear
arrangements in atomic waveguide structures \cite{Sinha:05}.

While vortices are interesting from a fundamental point of view, it
has recently been pointed out that the winding number of a single
vortex can be engineered to create a topologically stable quantum bit
for applications in quantum information
\cite{Kapale:05,Thanvanthri:08}.  Hallwood {\it et al.} have shown
that to create proper macroscopic superposition states of angular
momentum one needs to identify states which are close in energy,
strongly coupling, and, at the same time, well separated from any other
states \cite{Hallwood:07}. While these are difficult conditions to
fulfill in gaseous Bose-Einstein condensates with short range
interactions, the promise of long-lived qubits makes the effort
worthwhile.

A second fundamental building block of quantum information processing
is the ability to measure and control the interaction between two
qubits. While interactions between vortices can be studied in
Abrikosov lattices, it would be of advantage to find a geometry with a
lower number of nearest neighbours. Two recent works have shown that
systems where each vortex has only two nearest neighbours can be
created in atomic waveguide structure \cite{Sinha:05} or in
anisotropic, harmonic traps in the limit of weak interactions
\cite{Oktel:04}.

Here we extend the two works above by considering a Bose-Einstein
condensate deep in the non-linear regime (Thomas Fermi limit) and
trapped in an anisotropic trap. By numerically determining the minimum
energy state, we find that for relatively small aspect ratios, $\lambda
\gtrapprox 2$, and moderate rotational frequencies the rotating
Abrikosov lattice is no longer the ground state of the condensate and,
instead, a linear crystal of vortices is formed along the soft axis of
the trap. Taking advantage of the symmetry of the system, we devise an
ansatz for a variational calculation in order to predict ground state
properties of the system. We investigate the critical stirring
frequencies needed to move from the $l=0$ to $l=1$ state and the
locations in the lattice for a small number of vortices of equal
rotational charge. Note that anisotropic traps in the strong rotation
limit have recently been investigated in \cite{Aftalion:09}.

The paper is organised as follows. In
Sec.~\ref{sec:VariationalCalculation} we introduce our variational
wavefunction and calculate the energy of the system as a function of
anisotropy. In Sec.~\ref{sect:oneA} we determine the critical stirring
frequencies for local and global stability of a single vortex in an
anisotropic trap and in Sec.~\ref{sect:EquiLattPos} we calculate the
ground state structure of a vortex crystal with a small number of
vortices. In Sec.~\ref{sec:VortexDynamics} we briefly discuss the dynamics of a linear vortex crystal in an anisotropic trap. In Sec.~\ref{sec:Conclusion} we conclude.

\section{Variational Calculation}
\label{sec:VariationalCalculation}

\subsection{Energy Functional}

We consider a condensate of $N$ atoms with atomic mass $m$ and
scattering length $a_{sc}$, trapped in a potential $U(\mathbf{r})$
that is rotating at frequency $\Omega$. The energy functional for such
a system is given by
\begin{align}
  \label{eq:Functional3D}
  E[\Psi^*,\Psi]=&\int d^2\mathbf{r}\ dz\;\Psi^*
        \left[-\frac{\hbar^2}{2m}\nabla^2+U(\mathbf{r})
              -\mathbf{\Omega}\cdot\mathbf{L}\right]\Psi \nonumber\\
   &+ \frac{N g_{3D}}{2}\int d^2\mathbf{r}\ dz\;|\Psi|^4\;,
\end{align}
where $\mathbf{L}$ is the angular momentum operator,
$g_{3D}=4\pi\hbar^2a_{sc}/m$ is the coupling constant and $\int
d^3\mathbf{r}\;|\Psi|^2=1$. Finding the equilibrium states for the above
functional in all generality is a formidable task and beyond our
analytical as well as numerical capabilities. We will, therefore, in the
following make some assumptions on the symmetry of our system.

First, we will consider rotation around one of the major axes
only. Therefore, by choosing the angular momentum operator to be
$L_z$, we can separate the wave function into the $xy$-plane and the
$z$-direction as $\Psi(x,y,z)=\psi(x,y) \phi(z)$.
Since the contributions to the energy functional from the kinetic,
potential and rotational energies from $\phi(z)$ are constant, we need
only consider the contributions made to the coupling constant from the
third dimension.  In the following we will therefore make the
assumption that the wave function in the $z$-direction is in the
Thomas-Fermi limit and rescale the coupling constant as $g_{2D} =
g_{3D}(m\omega_z/2\pi\hbar)^{1/2}$. This creates an effective,
two-dimensional Hamiltonian to work with. Specifying the anisotropy of
the harmonic trapping potential in the $xy$-plane by the parameter
$\lambda$, we get the following energy functional
\begin{align}
  \label{eq:Functional}
  E[\psi^*,\psi]=&\int d^2\mathbf{r}\;\psi^*
                  \Bigg[-\frac{\hbar^2}{2m}\nabla^2
                        +\frac{m\omega^2}{2}\left(x^2+\lambda^2y^2\right)
                  \nonumber\\ 
                 &\qquad\qquad\quad-\Omega L_z\Bigg]\psi \nonumber\\
  &+ \frac{N g_{2D}}{2}\int d^2\mathbf{r}|\psi|^4\;.
\end{align}

\subsection{Variational Ansatz}

In order to find a good variational ansatz we need to consider both
the effect the angular momentum has on the density of the condensate
and the contribution to the phase of the condensate. A convenient and
transparent way of doing this is to use the quantum hydrodynamic
form and split the wavefunction into its modulus and
phase~\cite{Fetter:01}
\begin{align}
  \label{eq:fullansatz}
  \psi(x,y) = |\psi| e^{iS}\;.
\end{align}
Considering the modulus first, we find that, in the Thomas-Fermi limit
of large particle numbers, a condensate carrying vortices can be
characterised by two length scales. The first is the overall
condensate size and is given by the Thomas-Fermi radius \cite{Baym:96}
\begin{align}
  \label{eq:tfradius}
  R = \sqrt{\frac{2\mu}{m\omega^2}}\;,
\end{align}
where $\mu$ is the chemical potential of the system incluing the
vortex lattice. The size of the vortex cores gives the second
characteristic length scale. For an isotropic trap this is given by
\cite{Castin:99}
\begin{align}
  \label{eq:healinglength}
  r_{core} \simeq \sqrt{\frac{\hbar^2}{m\mu}}\;.
\end{align}
and we will show later that this is still a valid expression in the
anisotropic case. Since with increasing particle number these
quantities are inversely proportional to each other, we can separate
our ansatz with respect to these two scales as
\begin{equation} 
  \psi=\psi_{TF}\psi_{VL}\;.
\end{equation}
The Thomas-Fermi part, $\psi_{TF}$, describes the background cloud and
is given by the well known expression
\begin{align}
  \label{eq:TF}
  \psi_{TF}=\left[\frac{1}{Ng}
            \left(\mu-\frac{m\omega^2}{2}(x^2+\lambda^2 y^2)
            \right)\right]^\frac{1}{2}\;.
\end{align}
The vortices are described by the function $\psi_{VL}$, which will
only deviate from unity close to the individual vortex cores and for
which we will use a product of $\tanh$ functions of variable width
$\kappa = 1/r_{core}$ \cite{Castin:99}. The full ansatz for the
wave-function modulus therefore reads
\begin{align}
  \label{eq:modansatz}
  |\psi|=\psi_{TF}\prod_{k=1}^n\tanh\left[\kappa|\vec{r}-\vec{a_k}
    R|\right]\;,
\end{align}
and describes $n$ vortices located at $\vec{a_k}$. Note that the
vortex positions are scaled in units of the Thomas-Fermi radius in the
soft direction $R$, which lets us restrict our values of $a_k$ to
between $0$ (trap centre) and $1$ (condensate edge in the soft
direction).  We also define $\mu_0$ to be the chemical potential of a
condensate with no vortices
\begin{align}
  \label{eq:mu0}
  \mu_0 = \sqrt{ \frac{m \omega^2 N g \lambda}{\pi}}\;.
\end{align}
\begin{figure}[t]
  \includegraphics[width=\linewidth]{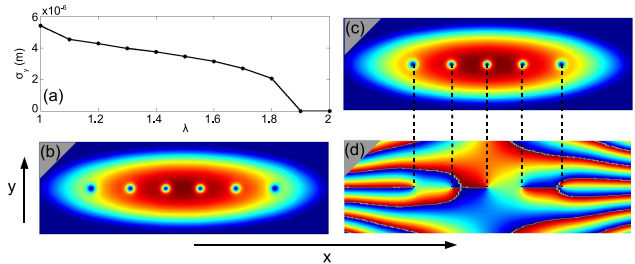}
  \caption{\label{fig:linear}Numerical simulations for a $^{87}$Rb
    gas of $N = 10^7$ atoms with $a_{sc}=4.67\times~10^{-9}m$ and trapping
    frequency $\omega = 20\pi s^{-1}$. (a) Standard deviation, $\sigma_y$, of the vortices from the x-axis for $1 \leq \lambda \leq 2$.  For $\lambda \geq 1.9$ the vortices form a linear crystal and $\sigma_y=0$. (b,c,d). Ground states for a condensate with parameters as above in a trap with aspect ratio of $\lambda = 2.5$. Each plot spans a range of
    $-36 \mu m < x < 36\mu m$ and $ - 11 \mu m < y < 11 \mu m $. (b) Ground state density for condensate rotating at $\Omega =
    0.7\omega$ showing 6 vortices.  (c)
    Ground state density for condensate rotating at $\Omega=0.65\omega$ showing
    5 vortices. (d) phase of the cloud shown in (c). The
    correspondence between the phase singularities and the density
    dips is indicated.     
    }
\end{figure}
To further reduce the number of free variables in the energy
functional, we will assume that we are in the anisotropic limit in
which the vortices are forming a linear lattice
\cite{Oktel:04,Sinha:05}. While this assumption at first glance looks
only reasonable for extremely anisotropic systems where the
Thomas-Fermi radius is of the order of the vortex core diameter, we
have found through numerical simulations that linear crystals are
already present for medium anisotropies if the number of vortices is
small. This can be qualitatively understood by realising that the
anisotropies lead to extended low density areas in the soft direction
and an alignment along this axis therefore minimizes the vortex self
energies as well as interactions. In Fig.~\ref{fig:linear}(a) we show
the standard deviation of the vortex core positions from $y=0$ for a
system of 5 vortices as a function of anisotropy. For an isotropic
trap the vortices form a triangular arrangement, which with increasing
values of $\lambda$ becomes squeezed along the tight direction. For
the relative moderate value of $\lambda=1.9$ the standard deviation
$\sigma_y$ goes to zero and remains there for all higher values of
anisotropy. Figs.~\ref{fig:linear}(b)-(d) show these numerically
determined ground states for an even and an odd number of vortices,
illustrating this linear geometry. Therefore we assume that
$\vec{a_k}=(a_k,0)$, i.e.~the vortices sit along the $x$-axis of the
trap. Furthermore, due to the symmetry of the system, we can assume
that for even numbers of vortices, vortex pairs are located at
$(a_k,0)$ and $(-a_k,0)$ and for odd numbers of vortices, a single
vortex will sit at the centre of the trap and the remaining vortices
will pair off as in the even case.

Finally, the phase itself can be broken up into the phase of the
condensate without vortices, $S_0$, and the phase from each vortex core
of charge $q_k$
\begin{align}
  \label{eq:totphase}
  S(x,y) = S_0(x,y) + \sum_{k=1}^n q_k \theta_k\;.
\end{align}
The first part describes the phase of the anisotropic condensate under
rotation without vortices and is given by
\cite{Castin:99,Svidzinsky:00}
\begin{align}
  \label{eq:phase1}
  S_0(x,y) = - \frac{m\Omega}{\hbar}\frac{\omega_x^2-\omega_y^2}
                                         {\omega_x^2+\omega_y^2}xy\;.
\end{align}
Since in our system the vortices are sitting along a single line along
which the phase is constant, we can neglect this contribution to the
phase and take
\begin{align}
  \label{eq:phase2}
  S = \sum_{k=1}^n q_k \theta_k.
\end{align} 

\subsection{Variational Energies}

Having the defined the ansatz and energy functional in the last
section, we can now calculate and minimise the different energies
involved. All our results for single and two vortex systems match up
with recent work by Castin and Dum \cite{Castin:99} in the isotropic
limit.

\subsubsection{Kinetic Energy}

Since we assume our system to be in the Thomas-Fermi limit, we will
neglect the spatial variation of the background condensate cloud when
calculating the kinetic energy. We can then rewrite $\tanh^2(x) =
1-\text{sech}^2(x)$ and take advantage of the fact that
$\text{sech}^2(x)$ is non-zero only within a small distance from the
vortex core. This allows us to neglect all terms that involve products
of two different vortex cores, i.e.~terms of the form
$\text{sech}^2(x_i)\text{sech}^2(x_j), (i\neq j)$ and leaves us the
following expression for the kinetic energy \cite{Castin:99}
\begin{align}
  \label{eq:ekin}
  E_{\text{kin}} =&
  \frac{\hbar^2}{2m}\int d^3\mathbf{r}\;|\psi_{TF}|^2 
  \sum_{k=1}^n\tanh^2\left[\kappa|\vec{r}-\vec{a_k}R|\right](\nabla\theta_k)^2
  \nonumber\\
  +&\frac{\hbar^2}{2m}\int d^3\mathbf{r}\; |\psi_{TF}|^2 
  \sum_{k=1}^n\kappa^2(\tanh \left[\kappa|\vec{r}-\vec{a_k} R|\right])^2 
  \nonumber\\
  +&\frac{\hbar^2}{2m}\int d^3\mathbf{r}\; |\psi_{TF}|^2 \sum_{k=1}^n
  \sum_{k'\neq k} q_k q_{k'} \nabla \theta_k \cdot \nabla \theta_{k'}\;.
\end{align}
The first two terms describe contributions from single vortices and
the last one accounts for the contributions from two vortices. As
expected from our ansatz, the anisotropy is completely contained in
the Thomas-Fermi parts. The single vortex terms can be
straightforwardly integrated to give
\begin{align}
  E_{\text{kin}_1}=&\sum_{k=1}^n \frac{\hbar^2\omega^2}{\mu_0}(1-a_k^2) 
     \left(\ln(\kappa^2 R^2)+\frac{1}{2}\ln(1-a_k^2)-c\right) \nonumber\\
   &-\sum_{k=1}^n \frac{\hbar^2\omega^2}{\mu_0} 
                 \frac{1-a_k^2(-5+\lambda)+\lambda}{2(1+\lambda)}\;,
\end{align}
where $c = (4\ln(2)-1)/6$ is a constant. The remaining term,
$E_{\text{kin}_2}$, cannot be integrated analytically and we will
solve it numerically when discussing interacting vortex systems below.

\subsubsection{ Potential and Rotational Energies}

The potential and rotational energies are given by
\begin{align}
  \label{eq:rotpotint}
  E_{\text{pot}} =& \sum_{k=1}^n\int d^2\mathbf{r}\;|\psi|^2 
    \left[\frac{m\omega^2}{2}(x^2+\lambda^2y^2)+\frac{Ng}{2}|\psi|^2\right]\;,\\
  E_{\text{rot}} =& -\sum_{k=1}^n\hbar\Omega\int d^2\mathbf{r}\;|\psi_{TF}|^2 
   \left( x\frac{\partial}{\partial y} - y\frac{\partial}{\partial x} \right) 
   q_k \theta_k \;,
\end{align}
and can be integrated to give
\begin{align}
  \label{eq:rotpot}
E_{\text{pot}}=& \sum_{k=1}^n2\mu_0 c \frac{(1-a_k^2)}{\kappa^2R^2}\;,\\
E_{\text{rot}}=&-\sum_{k=1}^n\frac{q\hbar\Omega}{2}\frac{\lambda^2(3-4a_k^2)-1}
              {\lambda^3}\;.
\end{align}
To be able to calculate the rotational integral we have made one
further approximation by considering only the area traced out by a
circle around the vortex whose radius is the same size as the trap in
the tight direction. This means that we only take into account the
cylindrically symmetric contribution to the phase close to the vortex
and neglect the asymmetric contributions from the low-density edges of
the cloud~\cite{Svidzinsky:00}.

The total energy of the system can now be expressed as
\begin{align}
 \label{eq:energytot}
 E=\frac{2\mu_0}{3}+\sum_{k=1}^n W(a_k)+\sum_{k=1}^n\sum_{k'>k}^n I(a_k,a_{k'})\;,
\end{align}
where the single vortex self energy is given by
\begin{align}
 \label{eq:W1}
 W(a)=&E_{\text{kin}_1} + E_{\text{pot}} + E_{\text{rot}}\nonumber\\ 
     =&\frac{\hbar^2\omega^2}{\mu_0}(1-a^2)
       \left(\ln(\kappa^2 R^2)+\frac{1}{2}\ln(1-a^2)-c\right) 
       \nonumber\\  
      &-\frac{\hbar^2\omega^2}{\mu_0}
        \frac{1-a^2(\lambda-5)+\lambda}{2(1 + \lambda)} 
        \nonumber\\  
      &+2\mu_0 c \frac{(1-a^2)}{\kappa^2R^2}
        \nonumber\\  
      &+\frac{q\hbar\Omega}{2}
        \frac{1-\lambda^2(3-4a^2)}{\lambda^3}\;, 
\end{align}
and the interaction between two vortices of identical charge $q_k =
q_{k'}=\pm 1 (k\neq k')$ can be found from
\begin{align}
 \label{eq:I1}
 I(a_k,a_{k'}) = E_{\text{kin}_2} = 
  \frac{\hbar^2}{2m}\int d^2\mathbf{r}\;|\psi_{TF}|^2
  \nabla \theta_{\vec{a_k}} \cdot \nabla\theta_{\vec{a_{k'}}}\;.
\end{align}

\section{Critical Stirring Frequency for a Single Vortex}
\label{sect:oneA}

Let us first consider the critical frequency for a single vortex state
to become the stable ground state of the trap. Two factors have to be
taken into account, namely the existence of two different trapping
frequencies and the fact that the lower one sets an upper limit for
rotational stability in a harmonic potential. To study the physics of
a single vortex, we need only consider the vortex self energy, $W$. We
therefore minimise eq.~\eqref{eq:W1} first with respect to $\kappa$
and find that the size of a vortex core is related only to the
chemical potential of the cloud at the site of the vortex
\cite{Castin:99}
\begin{align}
 \label{eq:kr}
 (\kappa R)^2 = 4c(1-a^2)\frac{\mu_0^2}{\hbar^2\omega^2}\;.
\end{align}
In particular we note that this expression is independent of the
strength of the anisotropy which justifies our earlier assumption that
we can separate two length scales in the system. Using
eq.~\eqref{eq:kr} we can then express the self energy as
\begin{align}
 \label{eq:S2}
 W =& \frac{\hbar^2\omega^2}{\mu_0}(1-a^2)
      \left(\ln\left(\frac{2\sqrt{c}\mu_0}{\hbar\omega}\right)+
      \frac{1}{2}\ln(1-a^2)+c\right) 
      \nonumber\\
    & +\frac{\hbar^2\omega^2}{\mu_0} 
      \left(3a^2(\lambda^2-5)- \frac{3a^2}{1+\lambda} \right)
      \nonumber\\  
    & +\frac{q\hbar\Omega}{2}\frac{1-\lambda^2(3-4a^2)}{\lambda^3} 
\end{align}
where again the first line corresponds to the kinetic energy, the
second corresponds to the potential energy, the third corresponds to
the rotational energy. From this expression for the energy, we can
derive the two critical stirring frequencies, $\Omega_l$ and
$\Omega_g$, corresponding to the appearance of a local and global
energy minimum, respectively.

The local minimum corresponds to the point where $W$ changes from
having a local maximum to a local minimum at the centre of the trap
and is found by setting $\partial_a^2 W = 0$ at $a=0$
\begin{align}
 \label{eq:local}
  \Omega_l=\frac{\hbar\omega^2}{\mu_0}\frac{2\lambda^3}{3\lambda^2-1}
           \ln\left(\frac{C\mu_0}{\hbar\omega}\right)\;,
\end{align}
where $C = 2\sqrt{c} e^{c}$. For this value of the stirring frequency
a vortex becomes locally stable in the condensate.

The global minimum is the frequency for which which the energy of the
condensate with a vortex at the centre has lower energy than the
condensate without the vortex, i.e.~$W=0$ at $a=0$
\begin{align}
 \label{eq:global}
 \Omega_g=\frac{\hbar\omega^2}{\mu_0}\frac{\lambda}{2}
          \left[\frac{4+\lambda}{1+\lambda} 
          + \ln\left(\frac{C\mu_0}{\hbar\omega}\right)\right]\;.
\end{align}
\begin{figure}[tb]
   \includegraphics[width=\linewidth]{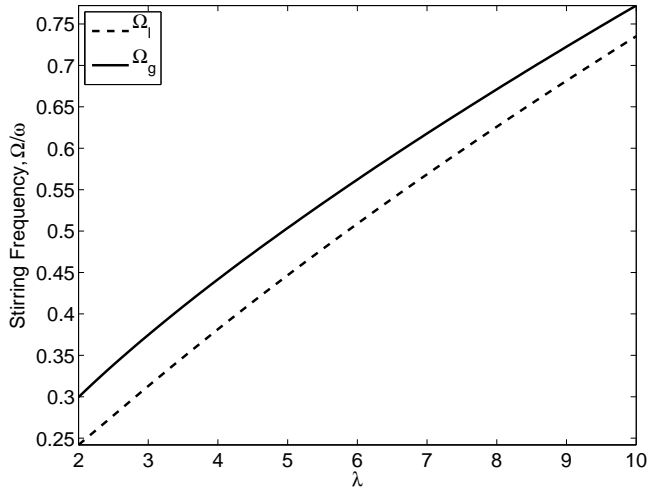}
   \caption{\label{fig:omegacrit} Critical frequency for a single
     vortex as a function of aspect ratio, $\lambda$. Stirring
     frequencies are defined in terms of the trapping frequency in the
     soft direction, $\omega$. For stirring frequencies greater than
     $\omega$, the cloud becomes unstable. $N = 1\times~10^5$, $\omega = 20\pi
    s^{-1}$, and $a_{sc}=4.67\times~10^{-9}m$ is the s-wave scattering
    lenghth for Rubidium-87.}
\end{figure}
Both quantities are shown in Fig.~\ref{fig:omegacrit} as a function of
anisotropy. One can see that both expressions above are not simple
extensions of the isotropic version, as they do not scale linearly
with $\lambda$. This agrees with earlier results showing the
appearance of irrotational velocity fields in rotating anisotropic
traps before vortex nucleation \cite{Svidzinsky:00}.

\section{Equilibrium lattice positions}
\label{sect:EquiLattPos}

In the following we will calculate the equilibrium positions of small
numbers of vortices in anisotropic traps as a function of
anisotropy. While for symmetry reasons a stable single vortex will sit
in the centre of the harmonic trap, the positions of larger numbers of
vortices are determined by a complex interplay between the single
vortex energies of eq.~\eqref{eq:W1}, and the vortex-vortex
interactions given by eq.~\eqref{eq:I1}. As argued above, we will
assume that the vortices are arranged in a linear crystal which is
aligned along the $x$-axis and symmetric with respect to the $y$-axis
of the trap.

\begin{figure}[tb]
  \includegraphics[width=\linewidth]{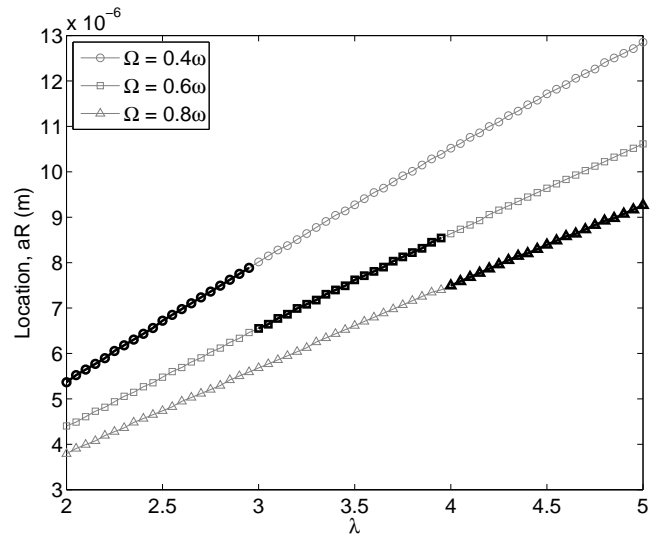}
  \caption{\label{fig:two_vort_lambda} Distance of each vortex in a
    vortex pair from the centre of the trap as a function of aspect
    ratio.  The results are shown for a gas of $N = 1\times~10^5$
    atoms of $^{87}$Rb. The trapping frequency is given by $\omega =
    20\pi s^{-1}$ and we display three different rotational
    frequencies $\Omega=0.4\omega, 0.6\omega$ and $0.8\omega$. The
    regions in which the two-vortex state is stable is marked by the
    black part of the graphs.}
\end{figure}

For two vortices of equal charge and located at $\pm a$ the integral
for the vortex-vortex interaction, eq.~\eqref{eq:I1}, can be
explicitly written out as
\begin{align}
 \label{eq:InterInt}
 I=\frac{\hbar^2}{2m}\int d^2\mathbf{r}\; 
   |\psi_{TF}|^2
   \frac{x^2+y^2-a^2}{\left((x-a)^2 +y^2\right)\left((x+a)^2 +y^2\right)}\;.
\end{align}
This expression cannot be fully integrated analytically and we
therefore minimise the complete energy functional numerical with
respect to the vortex positions $\pm a$. The results are shown for
different rotational frequencies in Fig.~\ref{fig:two_vort_lambda} as
a a function of anisotropy. The regions in which the two-vortex
molecule is the stable ground state of the system are marked by the
black parts of the curve. As the above minimisation procedure does not
give any information about the stability of the system, we have
determined these areas through numerical ground state calculations.
Note that they are only indicative though, since pinning down exact
borders is a challenging numerical task beyond our capabilities. Also
note that we display real units here, since the Thomas-Fermi radius
changes as a function of the rotation frequency.

\begin{figure}[tb]
	\includegraphics[width=\linewidth]{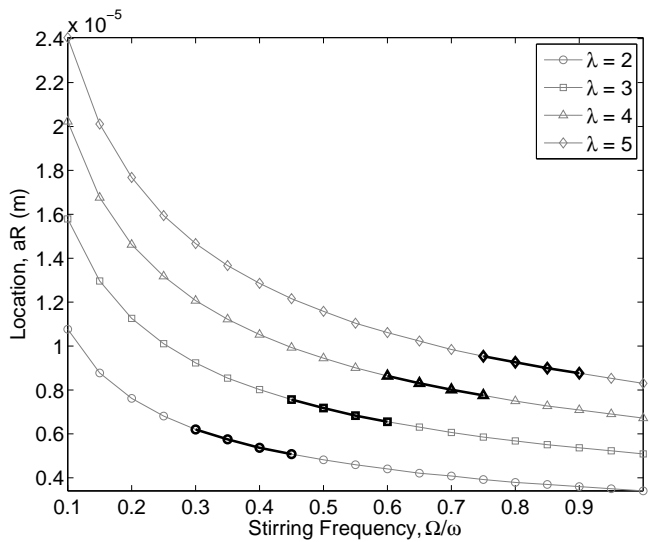}
	\caption{\label{fig:two_vort_Omega} Distance of each vortex in
          a vortex pair from the centre of the trap as a function of
          the stirring frequency. The system parameters are the same
          as in Fig.~\ref{fig:two_vort_lambda} and we display four
          different aspect ratios. The stable regions are again the
          parts marked by the black lines.}
\end{figure}

The first thing one can see from Fig.~\ref{fig:two_vort_lambda} is
that with increasing anisotropy the vortices are moving away from the
trap centre. This behaviour is a reaction to the increased
non-linearities due to the tighter potential in the transversal
direction, which has the effect of higher single vortex energies as
well as vortex-vortex interactions. Increasing the rotation frequency
has the opposite effect, as one can see from the relative position of
the three curves displayed. Faster rotation, i.e.~larger $\Omega$,
leads to an increased Thomas-Fermi radius in the soft direction and
allows for lower densities near the centre. The vortices therefore
move back towards the centre. The detailed form of this behaviour for
continuous $\Omega$ is shown in Fig.~\ref{fig:two_vort_Omega}.

\begin{figure}[tb]
  \includegraphics[width=\linewidth]{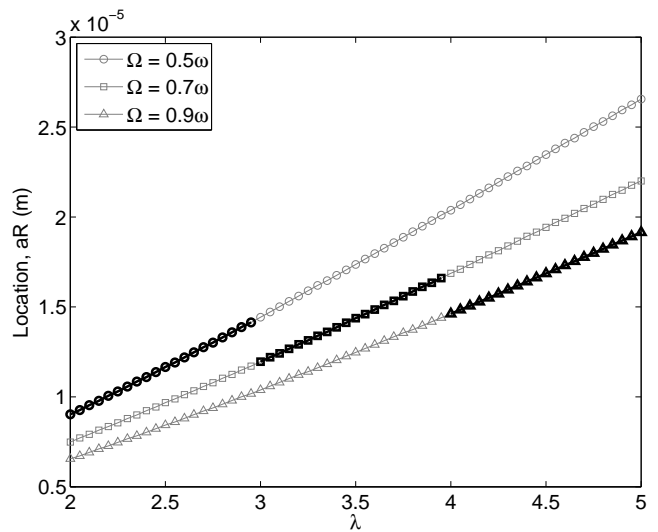}
  \caption{\label{fig:three_vs_lambda}Distance of outer vortices from
    central vortex in the three vortex case as a function of
    anisotropy for a range of stirring frequencies. The system
    parameters are the same as in the two vortex case.}
\end{figure}

\begin{figure}[tb]
  \includegraphics[width=\linewidth]{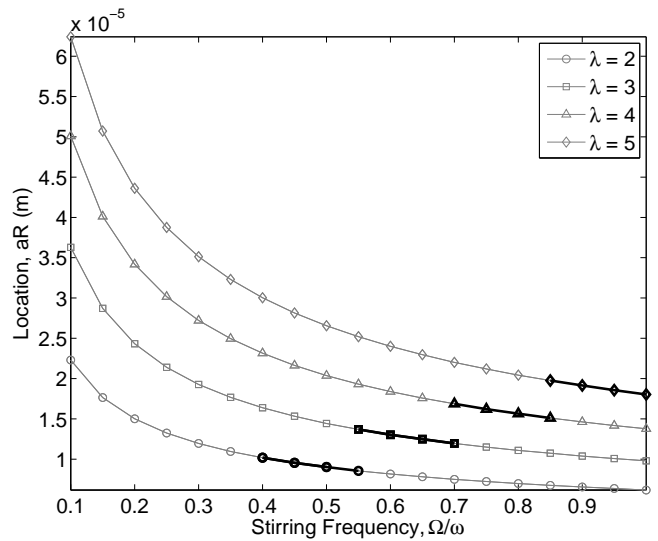}
  \caption{\label{fig:three_vs_Omega}Distance of outer vortices from
    central vortex in the three vortex case as a function of the
    stirring frequency for a range of anisotropies. The system
    parameters are the same as in the two vortex case.}
\end{figure}

For the three vortex case we again use the symmetry argument to choose
an ansatz in which the central vortex is located at $(0,0)$ and the
other vortices are located at $(\pm a,0)$. We now have include the
interaction between all three vortices, giving us three integrals, two
of which will be identical due to the
symmetry. Fig.~\ref{fig:three_vs_lambda} shows the location of the
outer vortices as a function of the aspect ratio. As with the two
vortex case, the vortices settle further away from the centre the more
anisotropic the trap is, however the distance is larger than in the
previous case. This is clearly due to the addition of the repulsive
centre vortex and when comparing the two and three vortex case, one
can see that the distance from the centre is about doubled. This
indicates that for at least small numbers of vortices within the TF
cloud the distance between two vortices is almost constant and which
confirms the overall trend that can be seen from the simulations shown
in Fig.~\ref{fig:linear}. This behaviour is also known from the
two-dimensional Abrikosov lattices \cite{Anglin:02}. Again, as in the
two vortex case and for the same reason, the vortices moves inward
with increasing trapping frequency (see
Fig.~\ref{fig:three_vs_Omega}).

It is easy to see how this method can be expanded for larger numbers
of vortices, taking advantage of the symmetry of the system. However,
the number of interaction terms required for $N$ vortices is
$N(N-1)/2$ requiring increased computational power and time.

\section{Vortex Dynamics}

While the above determines the ground state of the condensate, it does
not guarantee that the dynamical evolution maintains the linear shape
of the vortex crystal over time. In fact, the force on a vortex in an
inhomogeneous potential is directed perpendicular to the gradient of
the potential and one can therefore expect movement of each vortex
along an equipotential line of the trap. While in an isotropic trap
this would lead to a simple rotation of the linear crystal around the
centre of the trap, in an anisotropic trap this leads to a change in
relative distance between two vortices.  Shorter distances, however,
mean larger repulsive forces and one can see that beyond a critical
anisotropy the vortices will be unable to pass each other along the
perpendicular direction and will therefore maintain the linear crystal
structure.

\begin{figure}[tb]
  \includegraphics[width=\linewidth]{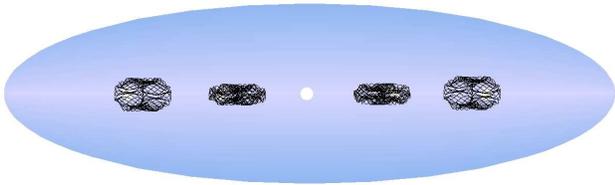}
  \caption{\label{fig:VortexDynamics} Trajectories of the single
    vortices over a time interval of $t=3.5$s. One can see that the
    outer vortices move around their equilibrium positions (white
    dots), but do not pass each other out. The central vortex remains stationary throughout. The system parameters are
    the same as in the figures above.}
\end{figure}

While the dynamical evolution is not the topic of the current work, we
have confirmed the above intuition by simulating the dynamical
behaviour of our system for a wide range of parameters and
anisotropies. In Fig.~\ref{fig:VortexDynamics} we show the
trajectories of five vortex cores in a trap with an anisotropy
parameter of $\lambda = 2.5$ for a duration of $t=3.5$s. One can see
that the vortices move around in the vicinity of their original
position, however they are not able to pass each other out. In fact,
the linear shape of the vortex crystal is maintained at any time and we
do not even observe any bending. At some point, for even longer
timescales, the backaction of the vortex movement will lead to
excitations of the background density and an exchange in vortex
position might happen. This will be the investigation of a future
work.

\section{Conclusion}
\label{sec:Conclusion}

Our work fills the gap between two recent investigations into vortex
lattice structures in anisotropic traps, one in the low density limit
\cite{Oktel:04} and the other in the fast rotational limit
\cite{Aftalion:09}. We have shown numerically that a small change in
the anisotropy of a moderately rotating anisotropic Bose-Einstein
condensate in the Thomas-Fermi limit changes the geometry of vortex
lattice from hexagonal to linear and we have presented a variational
analysis of such a system.  Using the simple symmetry of the linear
geometry we have devised a straightforward ansatz in the Thomas-Fermi
limit, which allowed us to split the energy function into single
vortex terms and two vortex interaction terms. From this we have been
able to determine the critical frequency for the creation of a single
vortex and found that as the aspect ratio increases, the critical
stirring frequency increases.

Minimising the energy functional for two and three vortex state, we
were able to determine the exact positions of the vortex cores. This
is a very tedious exercise for direct numerical treatments, as many
vortex lattices configurations have energies closely related to each
other. We found that the distance of the vortices from the centre
increases with increasing aspect ratio, however, it decreases as the
stirring frequency increases. At least for systems with small numbers
of vortices we found that the distance between neighbouring cores is
effectively constant.

Finally we briefly addressed the dynamical behaviour of a linear
vortex crystal and showed that for sufficiently anisotropic traps the
linear crystal geometry is maintained for very long time-scales.

\begin{acknowledgments}
  This project was supported by Science Foundation Ireland under
  project number 05/IN/I852.
\end{acknowledgments}

\end{document}